\documentclass[aip,graphicx]{revtex4-1}
\usepackage{graphicx}
\usepackage{siunitx}
\usepackage[english]{babel}
\usepackage{soul}
\usepackage{blindtext}
\usepackage{xcolor}
\usepackage{lineno}
\usepackage{amsmath}
\usepackage{multirow}
\usepackage[colorinlistoftodos]{todonotes}
\begin{document}

%\pagestyle{fancy}
%\rhead{\includegraphics[width=2.5cm]{vch-logo.png}}

%\linenumbers

%%%%%%%%%%%%%%%%%%%%%%%%%

\title{From Film to Flakes: Electronic Properties and Magnetization Variations in Yttrium Iron Garnet}

\author{Pia M. D\"uring}
\affiliation{Fachbereich Physik, Universit\"at Konstanz, 78457 Konstanz, Germany}

\author{Seema}
\affiliation{Fachbereich Physik, Universit\"at Konstanz, 78457 Konstanz, Germany}

\author{Roman Hartmann}
\affiliation{Fachbereich Physik, Universit\"at Konstanz, 78457 Konstanz, Germany}

\author{Sebastian Sailler}
\affiliation{Fachbereich Physik, Universit\"at Konstanz, 78457 Konstanz, Germany}

\author{Michaela Lammel}
\affiliation{Fachbereich Physik, Universit\"at Konstanz, 78457 Konstanz, Germany}

\author{Andrei Gloskovskii}
\affiliation{Photon Science, Deutsches Elektronen-Synchrotron DESY, 22607 Hamburg, Germany}

\author{Christoph Schlueter}
\affiliation{Photon Science, Deutsches Elektronen-Synchrotron DESY, 22607 Hamburg, Germany}

\author{Angelo Di Bernardo }
\affiliation{Fachbereich Physik, Universit\"at Konstanz, 78457 Konstanz, Germany}

\author{Sebastian T. B. Goennenwein}
\affiliation{Fachbereich Physik, Universit\"at Konstanz, 78457 Konstanz, Germany}

\author{Martina M\"uller}
\email{\\
\textbf{Corresponding author:} martina.mueller@uni-konstanz.de}
\affiliation{Fachbereich Physik, Universit\"at Konstanz, 78457 Konstanz, Germany}


\maketitle 

%Extended Abstract 
\section*{Abstract}
{\bf{
Yttrium iron garnet (YIG) is a ferrimagnetic insulator valued for its high Curie temperature, very low magnetic damping, and ability to support long-range spin-wave transport. These qualities have established it as a cornerstone material  in the field of spintronics and magnonics. Most studies on YIG so far have been focused on bulk crystals, thin films, and nanoparticles, including variants with substitutions at the yttrium or iron site. New morphologies such as sub-micron flakes have drawn interest recently as their geometry and mechanical flexibility might enable different device architectures.  However, detailed investigations combining their electronic structure and magnetic behavior remain scarce. In this work, we present a comparative study of the electronic and magnetic properties of a bulk-like YIG film and a sub-micron-sized YIG flake. Our results highlight the distinct behavior that emerges in sub-micron dimensions and point toward future uses for flake-based YIG in compact spintronics devices.}}

\newpage

\section{Introduction}

Yttrium iron garnet (Y$_3$Fe$_5$O$_{12}$, YIG) is a ferrimagnetic insulator known for its exceptional magnetic properties, including a high Curie temperature (560\,K), ultra-low Gilbert damping (6.7$\times$10$^{-5}$), and long spin-wave propagation length~\cite{1993_YIGsaga}. Its high Curie temperature ensures thermal stability across a broad temperature range, while the low magnetic damping makes it particularly suitable for microwave and magnonic devices~\cite{1993_YIGsaga}. These properties make YIG an ideal candidate for spintronic applications, especially in scenarios that require a magnetic material with insulating characteristics, making it one of the most highly sought-after materials in the field~\cite{Li2021, Dash2023}. Additionally, due to its non-metallic nature, YIG is routinely employed in spin pumping experiments~\cite{Haertinger2015}, where the absence of free charge carriers helps avoid electrical shunting and enables clean detection of spin currents. It also serves as a widely used magnetic insulating substrate for exploring magnetic proximity effects without introducing parasitic electrical conduction~\cite{Wang2015}. Collectively, these attributes have established YIG as a model material in the field of spintronics~\cite{Hirohata2020}.\\
Given its remarkable magnetic and optical properties, yttrium iron garnet (YIG) has been extensively investigated over the past decades in both bulk and thin film form, in its pure state as well as with various dopants~\cite{Arsad2023}. 
Substrate effects and Y-site (Fe-site) substitutions (e.g., Ce, Bi, Al, Ga) have been shown to significantly influence the magnetization, Faraday rotation, and Kerr effect in YIG~\cite{Ikesue2020, AsakerehRaad2020, Kehlberger2015, Dionne1970}. In the past few decades, pure and doped YIG nanoparticles have also been synthesized via various routes and explored for their application in frequency tunable microwave filters with magnetic field tuning~\cite{Sharma2016_YIGnp, Li2018_YIGnp}. More recently, a novel morphology of bulk YIG in the form of sub-micron flakes has been realized, offering promising properties such as shape anisotropy and mechanical transferability for the fabrication of device structures~\cite{2024_APL_Roman_YIGflakes}. Despite these advancements, comprehensive studies combining the electronic structure of such YIG flakes and related thin films remain limited.\\
This work presents a comparative analysis of the electronic and magnetic properties of a bulk-like pure YIG film and a sub-micron-sized YIG flake. Our findings provide a foundation for the future integration of YIG flakes into heterostructures via dry transfer techniques. These can be combined with other flakes or single-crystalline materials, including van der Waals systems, to realize novel spintronic devices, potentially extending into superconducting applications.

\section{Results}

\subsection{Methods} 

\subsubsection{Sample Preparation}

Many YIG flakes were mechanically exfoliated from a lab-grown Pb-free pure YIG single crystal as described by Hartmann et al.\,\cite{2024_APL_Roman_YIGflakes}.
The exfoliated YIG flakes were transferred onto an n-doped Si substrate with pre-patterned Au/Ti markers. However, the shape and orientation of the flakes could be random; therefore, several flakes had been exfoliated and tested for their chemical composition via Energy Dispersive X-ray Spectroscopy~\cite{2024_APL_Roman_YIGflakes}. 

A thin YIG film of 50\,nm was deposited at room temperature onto thermally oxidized Si wafers (Si/{SiO$_x$}) using RF sputtering from a YIG sinter target at \SI{2.7}{\cdot 10^{-3}\,mbar} argon pressure and \SI{80}{W} power, at a rate of \SI{0.0135}{nm/s}.
The post-annealing steps were carried out in a tube zone furnace under air~\cite{2024_PRB_Sebastian_YIGfilms}.
The Si substrate was chosen to prevent charging during the HAXPES experiments and to enable better comparison with the YIG flakes.

The Fe reference film on Si(100) was deposited using the dc magnetron sputtering technique in a vacuum chamber of base pressure of \SI{1}{\cdot 10^{-8}\,mbar} with well-defined calibration parameters with sputtering a 2 inch Fe target in working Ar pressure of \SI{1}{\cdot 10^{-4}\,mbar} at room temperature aiming for a thickness of 50\,nm. The resulting film is polycrystalline as expected. 

\subsubsection{Characterization}

To investigate the chemical composition of the deposited YIG film and the exfoliated YIG flake, we carried out Hard X-ray Photoelectron Spectroscopy (HAXPES) measurements at the P22 beamline of PETRA III (DESY, Hamburg)~\cite{Schlueter2019}.
Core level spectra of Fe 2p, Y 3p, and O 1s were recorded at photon energies of 2.8 keV and 6 keV, providing an information depth of about 11 nm and 24 nm, respectively~\cite{Sessa2017}. 
Therefore, HAXPES is more depth sensitive than conventional laboratory XPS (Al K$_\alpha$ or Mg K$_\alpha$), which typically has an information depth in the range of only few nanometers. 
A SPECS PHOIBOS 225HV electron analyzer was used at an emission angle of 2° and a pass energy of 30 eV, resulting in an overall energy resolution of approximately 300 meV~\cite{Mueller2022}. 
A schematic of the measurement setup is displayed in Fig.~\ref{fig:haxpes}(d).
To identify the YIG flake in the HAXPES instrument, the sample was systematically scanned at a fixed binding energy of 500 eV to find the Au markers as point of reference. 
The relative position of the YIG flake compared to the Au markers is known from optical microscope images.
After an Au marker was found, the sample was scanned at the fixed binding energy of Y 2p (E$_{Bin}$ = 2079 eV) until a signal from the flake was detected.
The lateral dimensions of the flakes vary between 5 - 20 ~\textmu m$^2$, therefore, the Si substrate is always measured as well using HAXPES with a spot size of 10 x 10 ~\textmu m$^2$, giving a large background contribution to the flake spectra. 
During the measurements, no charging of the sample was observed. 
From that, we can conclude that the nanoflake has a good conducting contact to the Si substrate, and therefore, the bonding procedure works well.\\
Magnetic hysteresis and domains in YIG film and flake were observed by longitudinal Kerr microscopy while simultaneously measuring hysteresis using a wide-field Kerr microscope by Evico Magnetics, Germany~\cite{Schafer2007}. Since the Kerr sensitivity in YIG is in blue light region, we used an LED-based light source to incident a light with a wavelength of 457\,nm at room temperature. 
The externally applied magnetic field was swept along the microscope in-plane sensitivity direction to plot an average intensity level of the whole area (film) or chosen region of interest (ROI) (in flake). We have measured the intensity component of the Kerr rotation parallel to the applied saturation field of 20\,mT.

\subsection{HAXPES}

\begin{figure*}[ht]
\centering
\includegraphics[width=170mm]{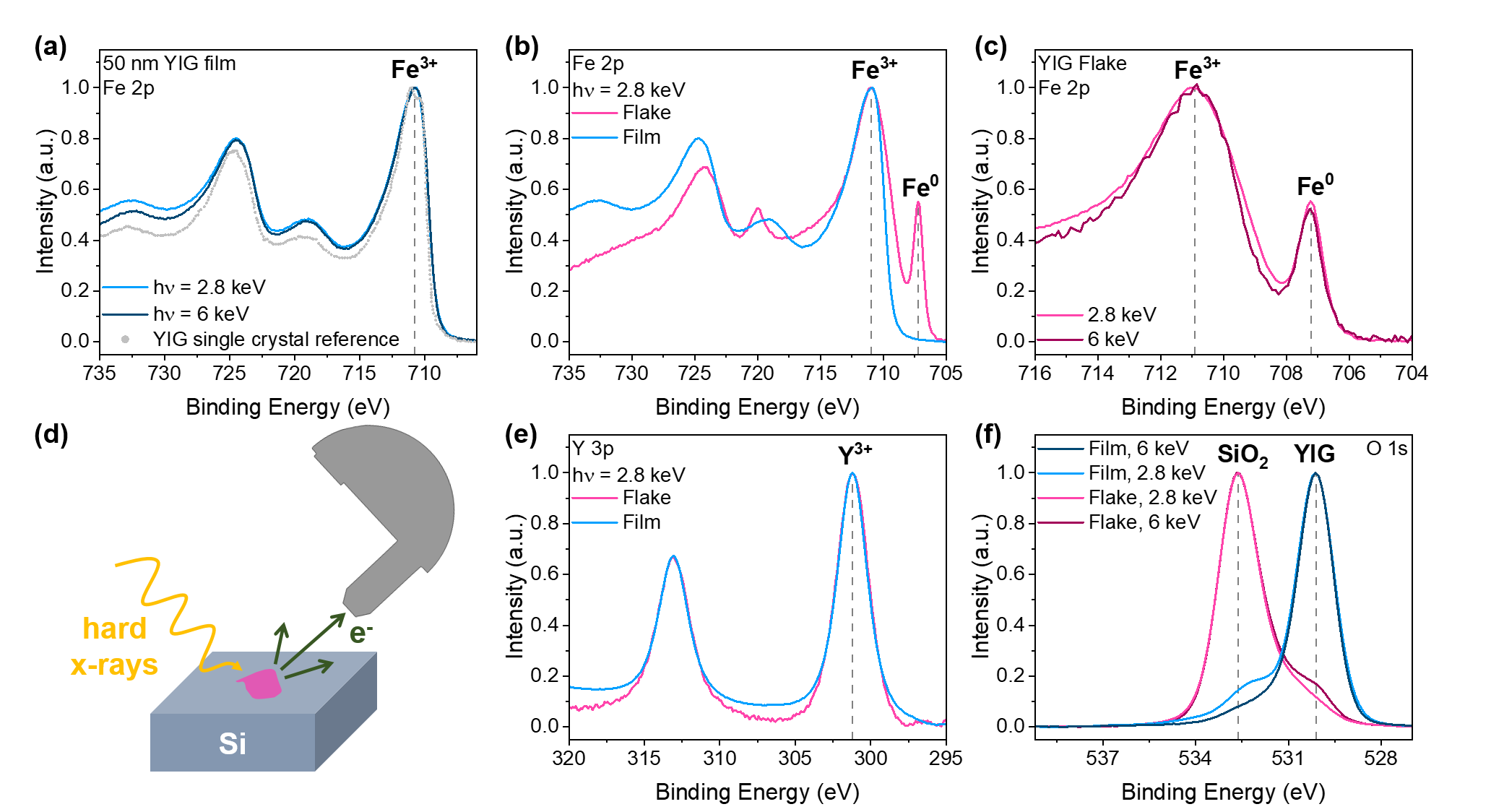}
\vspace{-8mm}
\caption{\label{fig:haxpes} HAXPES spectra of \textbf{(a) - (c)} Fe 2p core level, \textbf{(e)} Y 3p core level and \textbf{(f)} O 1s core level for the 50 nm YIG film (blue) and the YIG flake (pink) recorded at a photon energy of 2.8 keV and 6 keV. For a YIG single crystal, a reference~\cite{Chin2019} is shown in grey compared to the 50 nm YIG film in \textbf{(a)}. A schematic of the measurement setup is displayed in \textbf{(d)}.}
\end{figure*}

To compare the chemical composition of a 50 nm YIG bulk film and a YIG flake on a n-doped Si (100) substrate, Figure~\ref{fig:haxpes} displays the measured HAXPES spectra for the \textbf{(a) - (c)} Fe 2p core level, \textbf{(e)} Y 3p and \textbf{(f)} O 1s core level comparing the 50 nm YIG film (blue) and the YIG flake (pink). 
All spectra are normalized to the peak maximum. 

The analysis of the Fe 2p core level of the bulk YIG film in Fig.~\ref{fig:haxpes}(a) shows that the spectral shape of the thin YIG film does not change with the photon energy, which is a sign for a homogeneous deposited film. 
Compared to a YIG single crystal reference~\cite{Chin2019}, the HAXPES measurements confirm the stoichiometry of the synthesized YIG film with only Fe$^{3+}$ contribution (E$_{Bin}$ = 711 eV)~\cite{Hamed2020, Hamed2019} which is expected for a stoichiometric YIG sample. 
Although in the single crystal, the Fe 2p core level can be distinguished between the octa- and tetrahedral contribution of the Fe sites, this can not be expected for a thin film as can be seen in HAXPES measurements of a thin YIG film grown on GGG of Kobata et al.~\cite{Kobata2020}.
Therefore, the YIG film is well suited as a reference to the YIG nanoflakes. 

In Fig.~\ref{fig:haxpes}(b), the HAXPES spectra recorded at a photon energy of 2.8 keV for the Fe 2p peak of the YIG flake and the bulk film are compared.
For the flake, a clear Fe metal component (Fe$^0$) is present at a binding energy of 707.2 eV at the lower binding energy side of the Fe$^{3+}$ peak. 
Instead of only having a Fe$^{3+}$ contribution as it is expected for bulk YIG, the flake shows a metal peak with an approximate 2:1 intensity ratio (Fe$^{3+}$:Fe$^0$).
To determine a more accurate ratio, the Fe$^{3+}$ and Fe$^0$ peak areas need to be compared.
This would require a more detailed analysis, including background correction and peak fitting.

To distinguish whether the Fe metal component is located at the surface or is homogeneously distributed within the flake, we compare the Fe 2p$_{3/2}$ peak recorded at two photon energies with varying information depth in Fig.~\ref{fig:haxpes}(c).
At the photon energy of 2.8 keV, which is more surface sensitive, the intensity for the Fe$^{0}$ peak increases. 
This could be due to the small broadening of the Fe$^{3+}$ peak at 2.8 keV or to a slightly increased amount of Fe metal localized at the surface.
To verify the Fe$^{3+}$:Fe$^0$ ratio at both energies, and therefore the Fe metal distribution within the flake, a more detailed analysis is necessary to compare the ratio of the Fe$^{3+}$ and Fe$^0$ peak areas.

This leads to the question where the oxygen is transferred to if it is not bound to Fe anymore. 
Fig.~\ref{fig:haxpes}(e) displays the comparison of YIG bulk and YIG flake for Y 3p recorded at a photon energy of 2.8 keV where only a linear background was subtracted. 
For both spectra, the expected Y$^{3+}$ oxidation state is observed.
The peak shapes of the YIG film and the YIG flake seem similar without a new peak arising and any further background correction. 
For a complete analysis of the peaks, the background would have to be removed entirely which often introduces more error into the analysis.
Without any further background correction, Y does not appear to be affected by the oxidation change in Fe.
This still leaves us with the question where the oxygen is going to.

Hence, we compare the background corrected O 1s core level for the YIG film and the YIG flake recorded at 2.8 and 6 keV in Fig.~\ref{fig:haxpes}(f).
As mentioned before, HAXPES measurements at 2.8 keV are more surface sensitive than measurements at 6 keV. 
Additionally, the lateral size of the flake is smaller than the beam size spot. 
For that reason, a non-negligible part of the substrate is always measured simultaneously.
In all four spectra, we observe the oxygen component of YIG at a binding energy of 530.1 eV and of the Si substrate as SiO$_2$ at a binding energy of 532.6 eV. 
Furthermore, we see a higher amount of the SiO$_2$ substrate compared to the YIG component at 2.8 keV for the YIG flake (pink).
At 6 keV, we notice that the YIG ratio becomes more pronounced because we look deeper into the Si substrate and the SiO$_2$ contribution becomes smaller. 
The same effect is visible for the bulk YIG film (blue). 
At 2.8 keV, we observe a more pronounced SiO$_2$ contribution to the O 1s spectrum compared to the spectrum recorded at 6 keV. 
To discover differences in the O 1s core level, e.g. a oxygen deficiency, the substrate contribution would have to be removed from the spectrum, which introduced a large error into the analysis. 
Therefore, no reliable statement can be concluded from the O 1s core level.

Additionally to the HAXPES measurements, the elemental stoichiometry of the flakes was investigated using Energy Dispersive X-ray Spectroscopy (EDX). 
The measurements confirmed the elemental composition of the YIG flake, but EDX is not sensitive for light elements like O~\cite{Song2015} nor the oxidation state of the elements. %like HAXPES
Therefore, the YIG flake can still be oxygen deficient although it seems stoichiometric in EDX. 
Since a lower oxygen formation energy is needed to form Y$_2$O$_3$ compared to Fe and its oxides~\cite{Posadas2017}, it is likely that an oxygen deficient YIG flake would rather form Y$^{3+}$ than Fe$^{3+}$.
To maintain Y$^{3+}$ within the oxygen deficient flake, oxygen bonds with Fe could be broken.
This would explain the observed Fe metal (Fig.~\ref{fig:haxpes}(b)) and the stable state of Y$^{3+}$ (Fig.~\ref{fig:haxpes}(e)) in the YIG flake.

\subsection{Kerr Microscopy}

\begin{figure*}[ht]
\centering
\includegraphics[width=170mm]{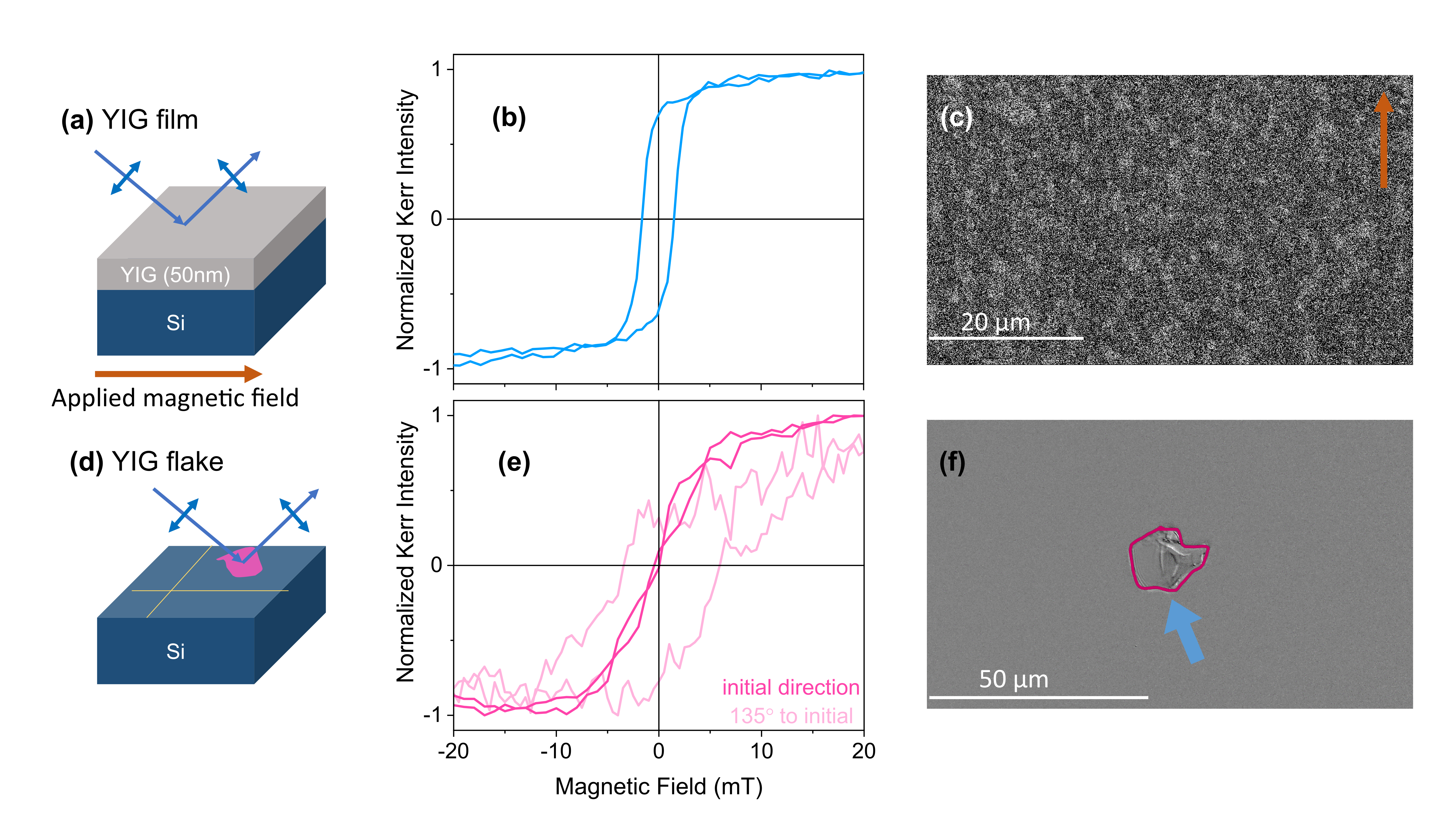}
%\vspace{-10mm}
\caption{\label{fig:fig2} \textbf{(a, d)} Schematic showing a YIG film and YIG flake measured in longitudinal geometry using blue light. \textbf{(b, e)} show measured hysteresis loops for the YIG film and flake. The arrow in \textbf{(c)} shows the applied field direction and Kerr sensitivity. The flake was also measured by rotating it on its own axis and representative 135$^{\circ}$ w.r.t. its initial direction is  shown. \textbf{(c, f)} shows the domain pattern observed in YIG film and very weak in the flake. The arrows in \textbf{(c)} and \textbf{(f)} are a guide to the eye to highlight the differences in gray contrast with in ROI of the flake.}
\end{figure*}

The simultaneous measurement of hysteresis and domain images using the Kerr microscope has already been tested for EuO thin films and YIG flakes in recent studies~\cite{Seema2025, 2024_APL_Roman_YIGflakes}. Fig.~\ref{fig:fig2} shows schematics of the YIG film and flake measured in LMOKE geometry using a blue LED light source. The magnetic hysteresis loop obtained for the measured film is shown in Fig.~\ref{fig:fig2}~(b). A well-defined hysteresis with a coercive field of $\approx$ 1.7\,mT has been observed, which is slightly higher than values existing in the literature~\cite{Hauser2016, CaoVan2022, Akhtar2016}. Since the film is polycrystalline and has been subjected to an annealing process, such a coercive field could be expected. Fig.~\ref{fig:fig2}~(c) shows a domain image obtained near the coercive field for the YIG film. To distinguish the magnetic domains, the gray contrast was observed when the domains switch due to a change in the direction of the applied field. Here the difference in Kerr contrast of the oppositely aligned domain regions is not strong enough and clear boundaries of oppositely aligned domains could not be distinguished. This can be attributed to polycrystalline nature of the YIG film on Si and the low Kerr rotation angle (~0.01 to 0.1$^{\circ}$ for light of wavelength 400 to 700\,nm~\cite{Tomita2006}) for the ferrimagnetic YIG.

Similarly, for the YIG flake, as shown in Fig.~\ref{fig:fig2}~(e) and (f), a hysteresis and faded switching of domains could be observed in the central part of the flake (marked by an arrow). The Kerr contrast amplitude for the flake was even smaller than that of the film. The flake was also tested for shape anisotropy as observed by Hartmann et al.~\cite{2024_APL_Roman_YIGflakes}, and the observed hysteresis after rotation of the flake by 135$^{\circ}$ w.r.t. the initial direction showed a different loop with enhanced coercivity, as expected. This has been also observed in YIG nanoparticles~\cite{P2022_YIGnp}. The Kerr contrast amplitude values measured by our setup under identical settings have been tabulated in Table~\ref{table:table1}. Domain switching was also observed within the flake as can be seen in the multimedia file in the supplemental material~\cite{suppmat}. To compare the Kerr amplitudes for the YIG film with the flake, we took the Kerr signal out of a small region of interest (ROI) with the help of the software with an area similar to that of the flake. For comparison purposes, we also measured a pure 50\,nm Fe reference film on a Si substrate with the same ROI to check how the values of Kerr contrast amplitude vary for a room-temperature ferromagnetic film such as Fe under identical conditions in the setup. The observed Kerr amplitude, as well as some literature values needed for the comparison, have been tabulated in the Table \ref{table:table1}. The observed Kerr intensity amplitude for the YIG film and flake was found to be significantly less as compared to the reference Fe film.

\subsection{Modulating Fe$^{3+}$/Fe$^0$ in bulk YIG }

To understand the observations in the Kerr intensity amplitude and the presence of the Fe$^0$ peak in the HAXPES data, we propose a model based on substituting one-third of the Fe$^{3+}$ to Fe$^0$ in the bulk YIG material and calculating the effect on overall magnetization in YIG film and flake. Such substitutional Fe-site doping of YIG and increased magnetization has been tested with Ni doping by Puspitasari et al.~\cite{P2022_YIGnp}. For these calculations, we assume that the YIG lattice does not undergo any lattice distortions, unlike when the Y-site gets substituted in YIG. We calculated the number of Fe atoms contributing to the magnetization in YIG in a specific volume. Since we aim to correlate these calculations to Kerr microscopy results, we consider this volume to be the same volume contributing to the Kerr amplitude in our setup. With a 50x magnification lens, the approximate area under observation for the flake is 100~\textmu m$^2$. The information depth of MOKE in most metals is 25\,nm, while YIG being an insulator it can be much higher due to low light absorption~\cite{book2006magnetism}. However, to simplify the calculations, we consider it 20\,nm. This gives us the volume under observation (V) to be 2$\times$10$^{-18}$~m$^3$. Using the basic solid-state physics relation below, we calculated the number of YIG formula units (N$_{YIG}$) as: 
\begin{equation}
\mathrm{N_{YIG}}=\frac{\rho_{\text {YIG }} \times \text { volume }}{\text { molar mass }} \times N_{A}
\end{equation}
where $\rho$$_{YIG}$ is the density of YIG which is 5.17\,g/cm$^3$ ~\cite{densityYIG} , the molar mass of YIG is 0.73\,kg/mol, and N$_A$ is Avagadro's number (6.022 $\times$ 10$^{23}$\,mol$^{-1}$). This accounts to 8.45$\times$ 10$^{9}$ YIG formula units and 4.22$\times$ 10$^{10}$ Fe atoms. Since each YIG unit cell has 5 Fe$^{3+}$ ions sitting at either tetrahedral or octahedral positions, they contribute different magnetic moment antiferromagnetically to the net magnetic moment of YIG~\cite{Bouguerra2007, Dash2023}. Literature on magnetic exchange interactions in YIG emphasizes the robust local spin magnetic moments of Fe, highlighting the significance of Fe-Fe interactions in determining YIG's magnetic properties~\cite{Gorbatov2021}. Considering the contributions from 2 octahedral Fe$^{3+}$ ions to be 3.75\,$\mu$$_B$ and from 3 tetrahedral Fe$^{3+}$ ions to be 3.70\,$\mu$$_B$, the overall magnetic moment from one YIG molecule comes out to be 3.60\,$\mu$$_B$, which matches the values observed in literature~\cite{REgarnetBook, Bouguerra2007}. Scaling this value to assumed V, we get an overall magnetization from YIG to be -3.25$\times$10$^{10}$~$\mu$$_B$ or 30.16$\times$10$^{-14}$~A/m$^2$ arising from 4.22$\times$10$^{10}$ Fe atoms in an ideal and pure YIG sample (using 1 Bohr magneton ($\mu$$_B$) = 9.27$\times$10$^{-24}$~A/m$^2$  ~\cite{bohrmagneton}). The negative sign here is here to show that the moments add up oppositely and one out of tetrahedral or octahedral site defines overall magnetization.

Now, as per our observed Fe$^0$ peak in HAXPES measurements, imagine a model of YIG where Fe$^0$ has substituted one-third of Fe$^{3+}$ atoms calculated above. We assume a one-third substitution depending on the ratio of Fe$^{3+}$ and Fe$^0$ peak intensities (2:1). Since we can not predict which octahedral or tetrahedral sites are going to be substituted, we assumed it to be distributed uniformly overall. For simplicity of calculations, we take the fraction of Fe$^0$ to be 30\% and the remaining 70\% to be Fe$^{3+}$. Therefore, the magnetization from the remaining 2.95$\times$10$^{10}$ Fe$^{3+}$ atoms in YIG comes out to be -2.26$\times$10$^{10}$~$\mu$$_B$ and contribution from 1.267$\times$10$^{10}$ Fe$^0$ atoms to be 2.78$\times$10$^{10}$~$\mu$$_B$, assuming each Fe$^0$ has a magnetization of 2.2~$\mu$$_B$ as in pure Fe metal~\cite{blundell2001magnetism}. Summing up, the magnetization in Fe$^0$-rich YIG would become 0.51$\times$10$^{10}$~$\mu$$_B$ or 4.75$\times$10$^{-14}$~A/m$^2$, which is significantly less than the magnetization of pure YIG as calculated above.

For reference, we calculated the number of Fe atoms in the same volume V in a Fe film. Taking the density of Fe to be 7.87~g/cm$^3$ ~\cite{arblaster2018}, the number of atoms in the same V is 1.69$\times$ 10$^{11}$. Assuming, as already mentioned, that each Fe$^0$ has a magnetization of 2.2~$\mu$$_B$ as, in pure Fe metal, the magnetization from all Fe atoms in V comes out to be 3.47$\times$10$^{-12}$~A/m$^2$. These values have also been tabulated in Table~\ref{table:table1}. The calculated values of YIG film as well as flake are much less as compared to the pure Fe film as expected. 

\subsection{Correlating model and experiments}

\begin{table}[ht]
\begin{center}
\caption{The theoretical values of magnetization per Fe atom of bulk YIG and Fe film from literature has been tabulated along with the calculated magnetization values. For comparison, the Kerr amplitude from the contrast as observed in the YIG film and flake has been compared with reference Fe film.}
\label{table:table1}
\begin{tabular}{|c|c|c|c|c|} 
 \hline
Sample~&~YIG Bulk~&~YIG film~&~YIG flake~&~Fe film\\ 
 \hline
 Magnetization Theory ($\mu$$_B$/Fe atom)& 5~\cite{Bouguerra2007}, 3.6~\cite{REgarnetBook} & -- & -- & 2.2~\cite{blundell2001magnetism}\\ 
 \hline
 Magnetization in V by our calculation ($\times$10$^{-14}$~A/m$^2$)& -- & 30.16 & 4.75 & 347\\
 \hline
 Kerr contrast amplitude from experiment (a.u.)& -- &  8.34 & 2.01 & 18.98\\
 \hline
 
\end{tabular}
\end{center}
\end{table}

We correlate the calculated magnetization values with our Kerr microscopy measurements to understand the observed low Kerr contrast in the YIG flake. Given that the YIG flake originates from the bulk YIG crystal, one would expect its Kerr contrast amplitude and magnetization to be comparable to that of the bulk or at least the YIG film. However, as shown in Table~\ref{table:table1}, the Kerr contrast of the flake is approximately four times lower than that of the ideal YIG film. Since the Kerr contrast amplitude could be indirectly linked to the sample’s overall magnetization, this significant reduction suggests a decrease in net magnetization.

Our HAXPES measurements revealed that approximately one-third of the Fe atoms in the flake exist as elemental Fe$^0$. We attributed this observation to the surface reconstruction process occurring when a flake is exfoliated from the bulk crystal. Incorporating this into our model — by substituting one-third of Fe$^{3+}$ ions with Fe$^0$ — we observe a notable reduction in net magnetization. This is attributed to YIG’s compensated ferrimagnetic structure and the fact that Fe$^0$ carries a smaller magnetic moment than Fe$^{3+}$ in tetrahedral or octahedral coordination. Consequently, the reduced net magnetization in the flake explains the diminished Kerr contrast amplitude, as well as the less pronounced hysteresis and magnetic domain features, relative to the YIG film. These findings suggest that the ferrimagnetic nature of YIG is compensated in the flake morphology. To validate these findings further, a systematic study involving multiple flakes with varying crystallographic orientations and detailed electronic structure analysis is necessary.

\section*{Summary}

The study compares a 50\,nm sputtered YIG film with a mechanically exfoliated sub-micron YIG flake to explore how morphology affects electronic and magnetic properties. Its high Curie temperature (560\,K), low Gilbert damping, and long spin-wave propagation make it ideal for spintronics, yet submicron flakes introduce shape anisotropy and flexibility that remain less explored. Measuring HAXPES at 2.8\,keV and 6\,keV photon energies, the film showed only Fe$^{3+}$ signals, confirming the stoichiometry, while the flake exhibited an additional Fe$^0$ component more near the surface. Kerr microscopy revealed a strong hysteresis for the film but weaker magnetization and domain switching for the flake. Modeling a one-third Fe$^{3+}$-to-Fe$^0$ substitution in the YIG volume reproduced and explained the four-fold reduction in Kerr amplitude. The results show that exfoliation induces surface reconstruction and metallic Fe formation, reducing magnetic performance and underscoring the need for surface control in YIG-based spintronic devices.

\section*{Acknowlegements}
MM, PMD and S acknowledge funding by the Deutsche Forschungsgemeinschaft (DFG, German Research Foundation), SFB 1432 (Project B03) with Project-ID 425217212.
STBG, SS and ML acknowledge funding by the Deutsche Forschungsgemeinschaft (DFG, German Research Foundation), SFB 1432 (Project B01) with Project-ID 425217212.
RH acknowledges support from the DFG under project no. 493158779 as part of the collaborative research project SFB F 86 Q-M\&S funded by the Austrian Science Fund (FWF) project number LAP 8610-N.
ADB acknowledges funding from the Alexander von Humboldt Foundation in the framework of a Sofja Kovalevskaja grant.
The authors acknowledge DESY (Hamburg, Germany), a member of the Helmholtz Association HGF, for the provision of experimental facilities. 
Parts of this research were carried out at PETRA III using beamline P22.
Beamtime was allocated for proposal I-20211596.
Funding for the HAXPES instrument at beamline P22 by the Federal Ministry of Education and Research (BMBF) under contracts 05KS7UM1 and 05K10UMA with Universität Mainz, 05KS7WW3, 05K10WW1, and 05K13WW1, with Universität Würzburg is gratefully acknowledged.

\section*{Data Availability}

The data that support the findings of this study are available from the corresponding author upon reasonable request.

\bibliography{YIG.bib}

\end{document}